\documentclass[%
 reprint,
superscriptaddress,
 aps,
]{revtex4-2}

\usepackage{float}
\usepackage{amsmath}
\usepackage{array}
\usepackage{graphicx}
\usepackage{dcolumn}
\usepackage{bm}
\usepackage{xcolor}

\begin{document}

\title{Electronic structure of low-dimensional inorganic/organic interfaces: Hybrid density functional theory, $G_0W_0$, and electrostatic models}

\author{Jannis Krumland}
\email{jannis.krumland@physik.hu-berlin.de}
\affiliation{Physics Department and IRIS Adlershof, Humboldt-Universität zu Berlin, 12489 Berlin, Germany}
\author{Caterina Cocchi}
\email{caterina.cocchi@uni-oldenburg.de}
\affiliation{Physics Department and IRIS Adlershof, Humboldt-Universität zu Berlin, 12489 Berlin, Germany}
\affiliation{Institute of Physics, Carl von Ossietzky Universität Oldenburg, 26129 Oldenburg, Germany}

\date{\today}

\begin{abstract}
First-principles simulations of electronic properties of hybrid inorganic/organic interfaces are challenging, as common density-functional theory (DFT) approximations target specific material classes like bulk semiconductors or gas-phase molecules. Taking as a prototypical example anthracene physisorbed on monolayer MoS$_2$, we assess the ability of different \textit{ab initio} schemes to describe the electronic structure using semi-local and hybrid DFT. For the latter, an unconstrained three-parameter range-separation scheme is employed. Comparisons against many-body perturbation theory results indicate that DFT is substantially unable to make reliable predictions about interfacial properties. Hybrid functionals, while improving the accuracy of the MoS$_2$ band structure, do not systematically enhance the description of hybrid systems with respect to semi-local functionals. Neither approach provides a good starting point for $G_0W_0$, which, consequently, cannot provide much information beyond the correct energy level alignment. We show that non-empirically parametrized electrostatic screening models can accomplish the same task at negligible computational costs. Such schemes can include substrates of hybrid interfaces in good agreement with experimental data. Our results indicate that currently, fully atomistic, many-body simulations of weakly interacting hybrid systems are not worth the required computational resources. In contrast, ab-initio-parametrized effective models mimicking the environment offer a scalable alternative without compromising accuracy and predictivity.
\end{abstract}

\maketitle

\newpage

\section{Introduction}

Low-dimensional semiconductors and their interfaces with conjugated molecules have become systems of interest for designing next-generation optoelectronic applications~\cite{jari+16nl,zhan+18am,zhu+18sa}.
To this end, an accurate description of the electronic structure of these materials and reliable predictions of their level alignment are needed to choose optimal combinations of inorganic and organic components. 
First-principles methods based on density-functional theory (DFT) are the natural choice for these calculations, especially in the absence of experimental parameters that are necessary to set up model Hamiltonians~\cite{katz+23prb,thom+23ns}.

The predictive power of DFT is, however, partially limited by the approximations taken for the electronic exchange and correlation ($xc$).
Standard implementations based on the homogeneous electron gas model~\cite{kohnSham1965pr} are unreliable for inhomogeneous systems.
Recipes developed to overcome this limitation include the addition of a fraction of Hartree-Fock exact exchange~\cite{b3lyp}, giving rise to so-called hybrid functionals.
These methods substantially improve the performance of DFT in the prediction of the electronic structure of molecules~\cite{adam-baro99jcp} and, to some extent, of bulk materials~\cite{koll+13jpcm}.
In the latter, the Coulomb potential is mediated by a sizeable electronic screening that is not accounted for in Hartree-Fock theory.
This issue was partly cured by the introduction of so-called range-separated hybrid (RSH) functionals~\cite{savi-flad95ijqc}, which have been demonstrated to perform very well for organic materials~\cite{maro+08jcp,sini+11jctc,luft+14prb}, crystalline solids~\cite{bech+09pssb,gero+18jpcm}, and low-dimensional systems~\cite{jain+11prl}.

The success of this method stimulated its application also to hybrid interfaces composed of metal oxides decorated with molecules~\cite{mowb-migb16jctc,stah-rink17cp,hofm-rink17aelm}, for which it could deliver improved descriptions of level alignment and bandgaps with respect to standard DFT.
RSH functionals have been applied also to low-dimensional hybrid interfaces~\cite{shen-tao17ami,xie+19jpca,zhen+21jpcm,zhou+21aplm,krum-cocc21es,mela+22pccp,jing+14jmca}, where, however, their superiority with respect to cheaper, lower-level approximations cannot be claimed by the achieved results~\cite{jing+14jmca}.
The reason for this has been ascribed to the intrinsically different nature of organic and inorganic constituents and of the Coulomb interaction therein~\cite{drax+14acr}.

Many-body perturbation theory (MBPT) is considered the state-of-the-art method to describe the electronic structure of solids.
Originally formulated in the framework of Green's function theory~\cite{hedi1965pr}, this approach is now frequently applied on top of DFT to correct the deficiencies of the latter while retaining its unpaired predictive power. 
In particular, The $GW$ approximation for the electronic self-energy enables the inclusion of the quasi-particle (QP) correction to the DFT states, thereby leading to accurate estimates of band structures and bandgaps.
The price of this achievement is a much higher computational cost with respect to DFT, due to the need to evaluate nonlocal operators. While originally devised as a self-consistent method~\cite{hedi1965pr}, in practice, $GW$ calculations are often performed in a perturbative flavor ($G_0W_0$) to correct the electronic structure from DFT with a single iteration~\cite{hybe+loui1986prb}. This introduces a dependence on the DFT starting point based on the chosen approximation for the $xc$ functional. Attempts to go beyond this scheme and include some degree of self-consistency remain the exception, due to the higher numerical efforts entailed and the emergence of additional theoretical shortcomings~\cite{mart+2016book}.

In this work, we assess established first-principles methods such as DFT with either semi-local or range-separated hybrid functionals and $G_0W_0$ for the calculation of the electronic structure of a prototypical hybrid interface formed by anthracene adsorbed on a transition metal dichalcogenide (TMDC) monolayer. We thoroughly analyze the results provided by these approaches highlighting the respective strengths and weaknesses. These fully atomistic simulations are further contrasted against semi-empirical calculations in which substrates are approximated as macroscopic dielectric media. Such models are able to estimate the substrate-induced renormalization of the electronic structure of the TMDC and the adsorbed molecule. They are interfaced with DFT through \textsc{LayerPCM}~\cite{krum+21jcp}, a recently developed extension of the polarizable continuum model (PCM)~\cite{toma+05cr}. 
Corresponding results and numerical performance indicate these methods as reliable and cost-effective alternatives for the calculation of the electronic structure of hybrid interfaces.

This paper is organized as follows: In Sec.~\ref{sec:theory}, we review the theoretical methods adopted in this work, and in Sec.~\ref{sec:comput-details} we report the computational details.
In Sec.~\ref{sec:results}, we present the results of our study, starting from the analysis of the isolated constituents (Sec.~\ref{sec:constituents}) and moving on to the hybrid systems (Sec.~\ref{sec:hybrid}).
We discuss our findings in Sec.~\ref{sec:discussion} and we expose conclusions and outlook in Sec.~\ref{sec:conclu}.

\section{Theoretical background}
\label{sec:theory}

In this section, we review the methodologies adopted in this study. We recall the fundamental equations of DFT (Sec.~\ref{sec:DFT}) and MBPT (Sec.~\ref{sec:MBPT}), addressing the approximations that are relevant for the present work. Special attention is given to RSH functionals based on the Coulomb attenuation method (Sec.~\ref{sec:CAM}).

\subsection{Density functional theory}
\label{sec:DFT}

DFT~\cite{hohenbergKohn1964pr} is an established method to deal with the many-electron problem by mapping it onto a fictitious system of non-interacting particles in an effective potential~\cite{kohnSham1965pr}. These auxiliary particles are characterized by the orbitals $\phi_j$ and the eigenvalues $E_j$ obtained as the solutions of the Kohn-Sham (KS) equation:
\begin{align}\label{eq:ks}
\Bigl\{-\frac{\nabla^2}{2} + V_\mathrm{ext}(\textbf{r}) + &V_H[n](\textbf{r}) \nonumber\\&+ V_{xc}[n](\textbf{r})\Bigl\}\phi(\textbf{r}) 
= E\phi(\textbf{r}).
\end{align}
The effective potential in Eq.~\eqref{eq:ks} is spelled out in three terms. The interaction between the electrons and nuclei is accounted for by $V_\mathrm{ext}$. The Hartree potential, which is the major component of the electron-electron interaction, is known exactly as
\begin{align}\label{eq:ksham}
    V_H[n](\textbf{r}) = \int\mathrm{d}^3r'\,v(|\textbf{r} - \textbf{r}'|)n(\textbf{r}').
\end{align}
Here, $v$ is the Coulomb interaction and $n$ the electron density, calculated as
\begin{align}
    n(\textbf{r}) = \sum_j^\mathrm{occ}|\phi_j(\textbf{r})|^2, 
\end{align}
which depends on the occupied subset of KS orbitals $\phi_j$, thus turning the KS equation Eq.~\eqref{eq:ks} into a nonlinear one. The last term, $V_{xc}$, describing $xc$ effects, depends on the density, too, but its exact form is unknown and requires approximations. The simplest one, known as local density approximation and introduced directly in the seminal work by Kohn and Sham~\cite{kohnSham1965pr}, treats $xc$ as in the homogeneous electron gas.
Higher levels of approximations extend this concept by including gradients of the electron density as well as the kinetic-energy density as variables~\cite{burk12jcp}.
Alternative strategies for approximating $V_{xc}$ entail the mixing between the aforementioned schemes with some fractions of Hartree-Fock exact exchange, giving rise in \textit{hybrid functionals}.
These advances have contributed to the success of DFT in dealing with inhomogeneous systems such as molecules, clusters, and defective solids~\cite{bech16book}.

\subsection{Range-separated hybrid functionals}
\label{sec:CAM}

RSH functionals, first introduced by Andreas Savin in the 1990s~\cite{savi-flad95ijqc}, comprise a set of approximations for the exchange energy relying on the splitting of the Coulomb potential into a short- and a long-range part. 
Here, we focus on general triple-parameter RSH functionals based on the so-called Coulomb attenuation method (CAM)~\cite{cam-b3lyp}. The basic idea behind the CAM is the partition of the Coulomb interaction as 
\begin{align}\label{eq:cam}
    v(r) = [\alpha + \beta\,\mathrm{erf}(\gamma r)]&v(r)\nonumber\\ + &\left\{1-[\alpha + \beta\,\mathrm{erf}(\gamma r)]\right\}v(r),
\end{align}
where $\mathrm{erf}$ is the error function, and $\alpha$, $\beta$, and $\gamma$ are adjustable parameters. The exchange energy is linear with respect to the Coulomb interaction and thus similarly turns into the sum of two terms. One of these terms is approximated with semi-local DFT exchange, the other one with Fock exchange, calculated with the KS orbitals. Hence, we end up with a distance-dependent mixture of the two approximations to exchange, as opposed to the fixed fractions assumed in popular global hybrid functionals such as B3LYP~\cite{b3lyp}. Typically, the first term in Eq.~\eqref{eq:cam} is associated with Fock exchange, with $\alpha$ and $\alpha+\beta$ being the fractions of exact exchange for $r\rightarrow 0$ and $r\rightarrow \infty$, respectively, while the ``screening parameter" $\gamma$ controls the transition from one to the other. Throughout this work, we refer to the tuned hybrid functionals as ``CAM" to differentiate it from other common definitions of RSH, which usually reduce the dimensionality of the parameter space by imposing constraints. Here, we instead allow for full flexibility in order to render the quality of the results independent of the choice of such constraints. The representation of the mixed exchange operator in a plane-wave basis is a straightforward generalization of the formulas given in Refs.~\cite{gygi+bald1986prb} and \cite{brog+2009prb} for global ($\beta = 0$) and HSE-like ``screened" exchange ($\alpha+\beta= 0$), respectively (see Supporting Information, Sec.~S1).

\subsection{Quasiparticle correction from the $GW$ approximation}
\label{sec:MBPT}

A different avenue for taking on the many-electron problem is established through MBPT. QP states $\psi_j$ and generally complex-valued eigenvalues $E_j$ can be obtained by solving
\begin{align}\label{eq:quasiparticle}
\Bigl\{-\frac{\nabla^2}{2} + &V_\mathrm{ext}(\textbf{r}) + V_H[n](\textbf{r})\Bigl\}\psi(\textbf{r}) \nonumber\\&+ \int\mathrm{d}^3r'\,\Sigma_{xc}(\textbf{r}, \textbf{r}',E)\psi(\textbf{r}') 
= E\psi(\textbf{r}).
\end{align}
Eq.~\eqref{eq:quasiparticle}~\cite{hedi1965pr} bears similarity to the KS equation, Eq.~\eqref{eq:ks}, except that the local KS $xc$ potential $v_{xc}[n]$ is replaced by the non-local self-energy $\Sigma_{xc}$, which furthermore depends on the eigenvalue $E$. In this work, we concern ourselves only with the real part of the energy $E$, neglecting its imaginary part (inverse QP lifetime). While the KS orbitals $\phi_j$ resulting from Eq.~\eqref{eq:ks} are mere mathematical auxiliary contrivances for determining the 
electron density $n$, the QP states $\psi_j$ from Eq.~\eqref{eq:quasiparticle} actually correspond to environmentally dressed and almost independent particles, and their energies quantify electron removal and addition from occupied and to virtual orbitals, respectively. The electronic self-energy $\Sigma_{xc}$ is considered here in the $GW$ approximation,
\begin{align}\label{eq:self_energy}
    \Sigma_{xc}(\textbf{r}, \textbf{r}', E) \approx\frac{i}{2\pi} \int&\text dE'\,e^{-i0^+E'}\times\nonumber\\ \times&G(\textbf{r}, \textbf{r}', E')W(\textbf{r}, \textbf{r}', E-E'),
\end{align}
where $G$ is the single-particle Green's function and $W$ is the dynamically screened Coulomb interaction. This approximation neglects vertex corrections to the self-energy. 

For the atomistic calculations performed in this work, we consider several further approximations to Eq.~\eqref{eq:self_energy}. Discarding the vertex corrections already for $W$ corresponds to calculating the dielectric function in the so-called random-phase approximation (RPA)~\cite{hybe+loui1986prb}. It is explicitly computed at two frequencies calculations and consequently fitted with a plasmon-pole model to approximate the full frequency dependence~\cite{godb+need1989prl}. Finally, all Green's functions occurring in $\Sigma_{xc}$ and $W$ are parametrized with orbitals $\phi_j$ and energies $E_j$ from a DFT calculation and not updated self-consistently, leading to the single-shot $G_0W_0$ correction to the DFT-based electronic structure.

The CAM functional mentioned in Sec.~\ref{sec:CAM} departs from the original spirit of KS-DFT, in which independent particles reside in an effective \textit{local} potential; the admixture of Fock exchange instead results in a non-local $xc$ operator. Formally, this moves hybrid functionals away from DFT in the direction of the QP formalism, as the $xc$ self-energy acts in a similar way [Eq.~\eqref{eq:quasiparticle}]. However, there is another reason to view hybrid DFT from this perspective. A major difference between the original KS and the QP frameworks is the existence of the derivative discontinuity in the former, which prohibits the physical interpretation of unoccupied orbital energies as electron addition energies~\cite{sham+schl1983prl, perd+levy1983prl}. However, within generalized KS theory, which formally justifies the admixture of Fock exchange, it is possible to at least partially absorb the derivative discontinuity into the eigenvalue gap~\cite{seid+1996prb}. In this case, the virtual orbital energies potentially provide a significantly better estimate of electron affinities. Since these correspond to the eigenvalues of Eq.~\eqref{eq:quasiparticle}, the mixed functional can indeed be considered a static approximation to the $xc$ self-energy,
\begin{align}
    &\Sigma_{xc}(\textbf{r}, \textbf{r}', E) \approx -\left[\alpha+\beta\,\mathrm{erf}(\gamma|\textbf{r}-\textbf{r}'|)\right]v(|\textbf{r}-\textbf{r}'|)\rho(\textbf{r}, \textbf{r}')\nonumber\\
    &+\delta(\textbf{r}- \textbf{r}')\{ \nonumber\\
    &\hspace{0.8cm}(1-\alpha-\beta) V_x[n](\textbf{r})+\beta V_x^\mathrm{SR}(\gamma)[n](\textbf{r})+V_c[n](\textbf{r}) \nonumber\\&\hspace{0.2cm}\},
\end{align}
where $\rho$ is the density matrix and $V_x$, $V_x^\mathrm{SR}$, and $V_c$ are the exchange, short-range exchange~\cite{hse03, hse06} and correlation functionals, respectively (Sec.~S1). 

\subsection{Long-range screening: Model corrections}\label{sec:model}
The additional screening introduced by polarizable media in the vicinity of an electronic system can substantially change its QP energies. We define $\Delta W = W - W_0$, where $W$ is the screened interaction of the system as a whole and $W_0$ is that of one isolated subsystem. We consider hybrid interfaces partitioned into three subsystems: substrate, TMDC, and molecule.
We are mainly interested in the latter two, but we include the first one to mimic more closely experimental setups.
For the molecule, the TMDC makes the dominant contribution to $\Delta W$; for the TMDC itself, the substrate is most relevant. Other interactions turn out to be negligible, as shown below. 

In such scenarios of spatially separated subsystems, the polarization of the respective external media is dictated by the static limit of their dielectric response~\cite{inks1973jpc}. 
The QP correction within a single subsystem due to the additional screening by the external medium can be approximated as~\cite{neaton2006prl}
\begin{align}\label{eq:sigma_cl}
 \int\text d^3r\,&\text d^3r'\psi^*(\textbf{r})\Delta\Sigma_{xc}(\textbf{r}, \textbf{r}')\psi(\textbf{r}')\nonumber\\
 &\approx
\pm\frac 12\int\text d^3r\,\text d^3r'\,\Delta W(\textbf{r}, \textbf{r}')|\psi(\textbf{r})|^2|\psi(\textbf{r}')|^2,
\end{align}
where ``$+$" and ``$-$" apply to hole and electron orbitals, respectively. Eq.~\eqref{eq:sigma_cl} is directly applicable when focusing on the organic component, inserting for $\psi$ the molecular orbitals. 

In practice, we use an arguably more accurate procedure, including such polarization terms directly in the KS Hamiltonian through the PCM~\cite{toma+05cr}. To this end, the KS Hamiltonian of Eq.~\eqref{eq:ksham} is supplied with a polarization term, $V_\mathrm{PCM}$, representing the electrostatic potential due to environmental polarization obtained by computing a representative charge surface density enclosing the molecule~\cite{krum+21jcp}. By determining with such an enhanced Hamiltonian electron addition and removal energies through total (free) energy differences ($\Delta$SCF method), the classical self-energy due to external polarization [Eq.~\eqref{eq:sigma_cl}] is automatically featured in a self-consistent way, allowing for full relaxation of the electron density of the molecule. To calculate instead the effect of the substrate on the electronic structure of the TMDC and specifically the bandgap renormalization $\Delta E_g$, we set $|\psi|^2=\delta$ in Eq.~\eqref{eq:sigma_cl} and obtain~\cite{cho+berk2018prb}
 \begin{align}
    \Delta E_g = \Delta W(\textbf{0}, \textbf{0}) = \lim_{\textbf{r}\rightarrow\textbf{0}}\left[W(\textbf{r}, \textbf{0})-W_0(\textbf{r}, \textbf{0})\right].
\end{align}
 For all such calculations, the respective external media are modeled as macroscopic dielectrics characterized by their static dielectric constants.
 
\section{Computational Details}
\label{sec:comput-details}

DFT calculations for extended systems, including free-standing two-dimensional materials and their interfaces with molecular adsorbents, are performed with version~6.8 of the plane-wave-based \textsc{Quantum ESPRESSO} suite~\cite{qe2020}. This code has been locally modified to allow for the independent tunability of $\alpha$, $\beta$, and $\gamma$ in Eq.~\eqref{eq:cam}. The correctness of these changes has been verified by comparison with the results of the built-in implementations of the PBE~\cite{pbe}, PBE0~\cite{pbe0}, and HSE~\cite{hse03, hse06} functionals, which correspond to specific choices of these parameters.
Geometry optimizations are performed using PBE with a plane-wave cutoff for the wavefunctions at 60~Ry, including the pairwise Tkatschenko-Scheffler dispersion correction~\cite{ts} to capture the crucial van-der-Waals attraction between the molecule and the two-dimensional material. For ensuing single-point calculations with the relaxed structure, the wavefunction cutoff is lowered to 40~Ry. The $c$ lattice parameter of the hexagonal cell is set to 20~$\mathrm{\AA}$ in all cases, corresponding to a large vacuum layer between replicas in this (supposedly finite) direction. A dipole correction is included in the center of this vacuum layer. The lattice parameter $a$ of MoS$_2$ is determined as 3.18~\AA. For DFT calculations, we employ \textbf{k}-grids ($\textbf{q}$-grids) of $12\times 12\times 1$ ($4\times 4\times 1$) and $2\times2\times 1$ ($1\times 1\times 1$) for unit cell and supercell calculations, respectively [the $\textbf{q}$-grid samples the Brillouin zone in the calculation of the Fock exchange, see Eq.~(S3)]. The reciprocal lattice sum in the Fock exchange is cut off at a kinetic energy of 40~Ry. For hybrid-functional calculations of interfaces, a single-shot approach is used, computing the Fock exchange with PBE orbitals, then recalculating the density and the orbitals once with the mixed Hamiltonian, but not leading the exact-exchange superloop to full self-consistency. These three simplifications accelerate calculations with hybrid functional to a significant degree. The corresponding loss of accuracy is estimated to be around 0.1~eV, which is deemed sufficiently small to ensure the desired accuracy. 
Version 3.1 of the \textsc{Wannier90} code~\cite{pizz+2020jpcm} is deployed to map QP energies from \textbf{k}-grid-based calculations onto finely spaced $\textbf{k}$-paths by means of Wannier interpolation~\cite{marzari+2012rmp}.

The PBE electronic structure serves as the starting point for MBPT calculations in the RPA and in the $G_0W_0$ scheme to evaluate dielectric functions and QP corrections, respectively.
To this end, version 5.1 of the \textsc{Yambo} code~\cite{Yambo_2019} is used. The \textbf{k}-grids are enhanced to $24\times24\times1$ and $6\times6\times1$ in the unit cell and supercell, respectively. For the inverse dielectric function, the plasmon-pole approximation is assumed. A two-dimensional Coulomb cutoff~\cite{rozz+07prb} and the random-integration method using 10$^6$ points for all $\textbf{G}$-vectors with a kinetic energy below a cutoff of 40~eV are applied. The Bruneval-Gonze sum-over-states terminator is employed to accelerate convergence with respect to the number of empty bands~\cite{brun+gonz08prb}. $GW$ results are further converged by extrapolating them to a complete basis, \textit{i.e.}, to both an infinite number of conduction bands in the sum over states in both $W$ as well as $\Sigma_c$, and an infinitely large kinetic energy cutoff in the $\textbf{G}$-sums in $W$ (see Sec. S2). For free-standing MoS$_2$, calculations are performed with both 100 and 200 conduction bands, enabling the 
extrapolation with respect to the number of conduction bands, while the plane-wave cutoff is fixed for $W$ at 80~eV; a third calculation is conducted with 100 conduction bands and a cutoff of 103~eV for the extrapolation with respect to the plane-wave cutoff.
In the hybrid system, the same plane-wave cutoffs are employed while increasing the number of conduction bands to 659 and 859 for the corresponding extrapolation.

\textsc{LayerPCM}~\cite{krum+21jcp} as implemented in version 9.2 of the \textsc{Octopus} code~\cite{octopus2020} is employed in combination with the PBE functional for $\Delta$SCF calculations of the non-periodic molecule, with the TMDC and the substrate modeled as macroscopic dielectrics. A real-space grid is adopted, generated by sampling the union of all atom-centered spheres of radius 5~\AA\,with a spacing of 0.12~\AA. The PCM cavity is built in a similar fashion, interlocking spheres of radius 2.48~\AA\,on all C atoms. The cavity surface is approximated with 60 finite boundary elements per sphere, discarding those within the cavity or too close to one another. Simulations of the open-shell anionic and cationic molecules are performed without spin restriction. For model MoS$_2$, we use in-plane and out-of-plane dielectric constants of
$\varepsilon_\parallel = 16.85$ and
$\varepsilon_\perp = 12.30$, respectively, as well as a thickness of 
$t_1 = 5.35\,\mathrm{\AA}$ (see Fig.~S2). For the size $t_2$ of the vacuum layer between the TMDC and a potential substrate, we take a value of $t_2=0.95\,\mathrm{\AA}$. This set of parameters has been determined from first principles, as described in detail in Sec.~S3. 

 Scalar-relativistic and norm-conserving SG15 pseudopotentials~\cite{sg15_2015}, which are supported by all the adopted codes, are employed in all calculations. Supercell band structures are unfolded~\cite{boyk+klim2005prb, ku+2010prl, pred+zung2012prb, mayo+2020jcm, dirn+2021jpcc} to the primitive cell of MoS$_2$ with an in-house developed routine (Sec.~S4).

\section{Results}
\label{sec:results}

We present the results of this work starting from the analysis of the electronic structure of the constituents, molecule and TMDC, treated as isolated entities (Sec.~\ref{sec:constituents}). Subsequently, we discuss the proposed strategies to predict the properties of the hybrid interface (Sec.~\ref{sec:hybrid}).

\subsection{Subsystems}
\label{sec:constituents}

\subsubsection{Band structure of MoS$_2$: PBE and $GW$}

\begin{figure}[h]
    \centering
    \includegraphics[width=0.46\textwidth]{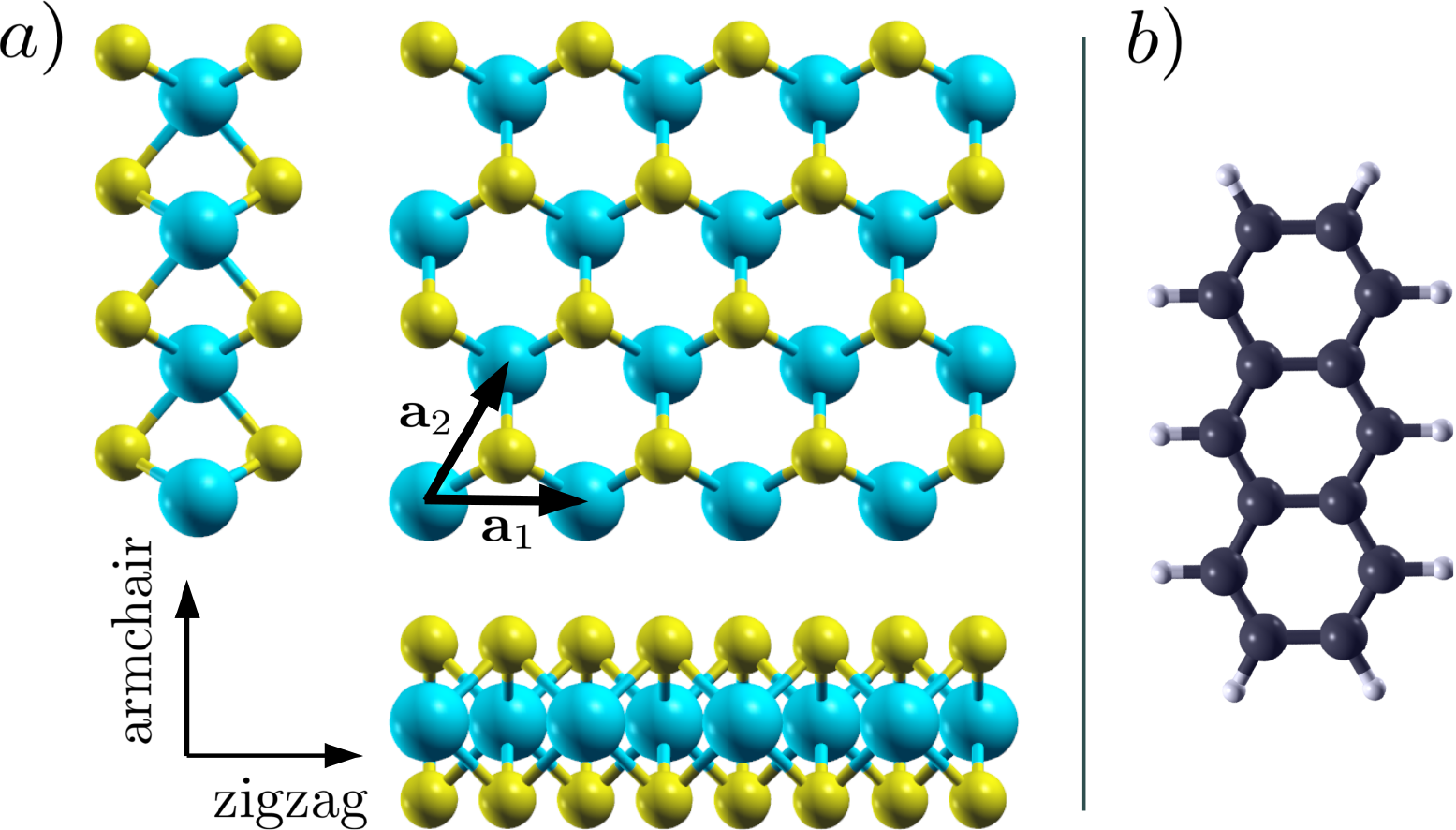}
    \caption{a) Single-layer hexagonal MoS$_2$, top and side view; $\textbf{a}_1$ and $\textbf{a}_2$ are the primitive cell vectors. b) Anthracene (ANT) molecule. Turquoise, yellow, grey, and white spheres represent Mo, S, C, and H atoms, respectively.}
    \label{fig:unit_geom}
\end{figure}

\begin{figure}[h]
    \centering
    \includegraphics[width=0.48\textwidth]{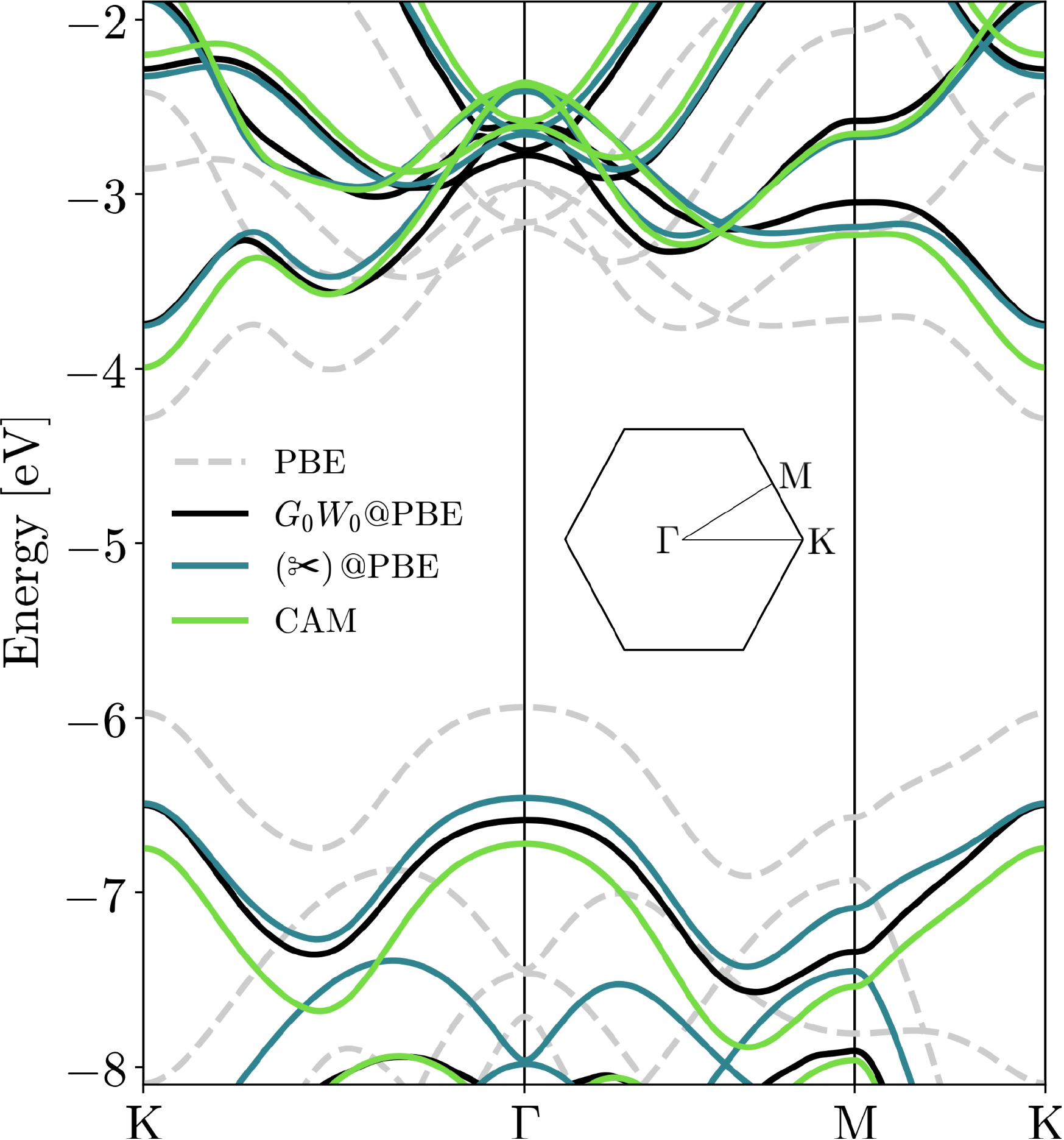}
    \caption{Band structure of monolayer MoS$_2$ calculated with the PBE functional, with $G_0W_0$@PBE, scissor-shifted PBE, and with a CAM functional tuned to reproduce the gaps at M and K, taking $\alpha = -\beta = 0.28$ and $\gamma = 0.034\,a_\mathrm{B}^{-1}$ (HSE-type range separation). The energy is offset with respect to the vacuum level. Inset: Brillouin zone of hexagonal 1L-MoS$_2$ with the high-symmetry points highlighted.}
    \label{fig:unit_bands}
\end{figure}

We investigate the band structure of a hexagonal 1L-MoS$_2$ [Fig.~\ref{fig:unit_geom}a)], examining the results obtained from PBE (Fig.~\ref{fig:unit_bands}).
The energy bands are plotted along paths connecting the high-symmetry points $\Gamma$, M, and K, which define the boundaries of the irreducible Brillouin zone of this material (inset of Fig.~\ref{fig:unit_bands}). M and K correspond to armchair and zigzag directions in real space, respectively [\textit{cf.} Fig.~\ref{fig:unit_geom}a)]. As established in the literature, 1L-MoS$_2$ has a direct bandgap at K which is responsible for its appealing optoelectronic characteristics~\cite{mak+10prl}.
Our PBE results reproduce this feature yielding a bandgap value at K of 1.69~eV. 
The QP correction obtained from $G_0W_0$ increases its magnitude by $\sim$1~eV, leading to the absolute value of 2.76~eV (Fig.~\ref{fig:unit_bands}).
This finding is within the range of $2.4-2.8$~eV (average $=2.67$~eV) reported in the literature based on the same method~\cite{ryou+2016sr, utam+2019ne, drue2017natcom, naik+jain2018prm, qiu+2015prl, qiu2016prb, sokl+2014apl, koms+kras2012prb, moli+2013prb, shi+2013prb, huse+2013prb, rama2012prb, conl+2013nl, chei+lamb2012prb, liang+2013apl}. Note that spin-orbit coupling is not taken into account in our calculation; it is expected to reduce the gap by some 0.1~eV by splitting the valence band maximum at K~\cite{rama2012prb}.

Absolute values for the band edges of MoS$_2$ are less frequently reported and converge much more slowly. The values of -6.50~eV and -3.74~eV found here for the highest valence band and the lowest conduction band, respectively, are in perfect agreement with the carefully converged $GW$ calculations reported in Ref.~\citenum{liang+2013apl}. The present results also confirm the observation made therein that the QP band-edge shifts are symmetric: the 1~eV of gap opening upon inclusion of the QP correction from $G_0W_0$ on top of PBE arises from a -0.5~eV shift of the valence band plus a +0.5~eV shift of the conduction band. This proves that absolute values for the band edges can in principle be obtained from only gap-converged $G_0W_0$@PBE calculations by aligning the mid-gap with that of the PBE starting point~\cite{toro+2011pccp}. 

In order to analyze more in-depth the many-body corrections introduced by the $G_0W_0$ approximation, we contrast the $G_0W_0$@PBE result with the one obtained by correcting the PBE band-structure with a scissors operator that aligns the valence band maximum (VBM) and the conduction band minimum (CBm) with those obtained from $G_0W_0$@PBE (Fig.~\ref{fig:unit_bands}).
From this comparison, we notice immediately that neither the upmost valence band nor the lowest conduction band coincides with those obtained from $G_0W_0$@PBE, once departing from the extrema at K. 
However, the scissor-shifted PBE result reproduces quite well the gap at $\Gamma$, albeit due to some cancellation of errors: both the highest valence states and the lowest conduction states are energetically overestimated by an equal amount of about $0.2$~eV. 
In contrast, large deviations from the $G_0W_0$@PBE result are found at M, where the QP correction features an additional gap opening influencing quite heavily the band dispersion between M and K.

\subsubsection{Band structure of MoS$_2$: CAM functional}

\begin{figure}[h]
    \centering
    \includegraphics[width=0.48\textwidth]{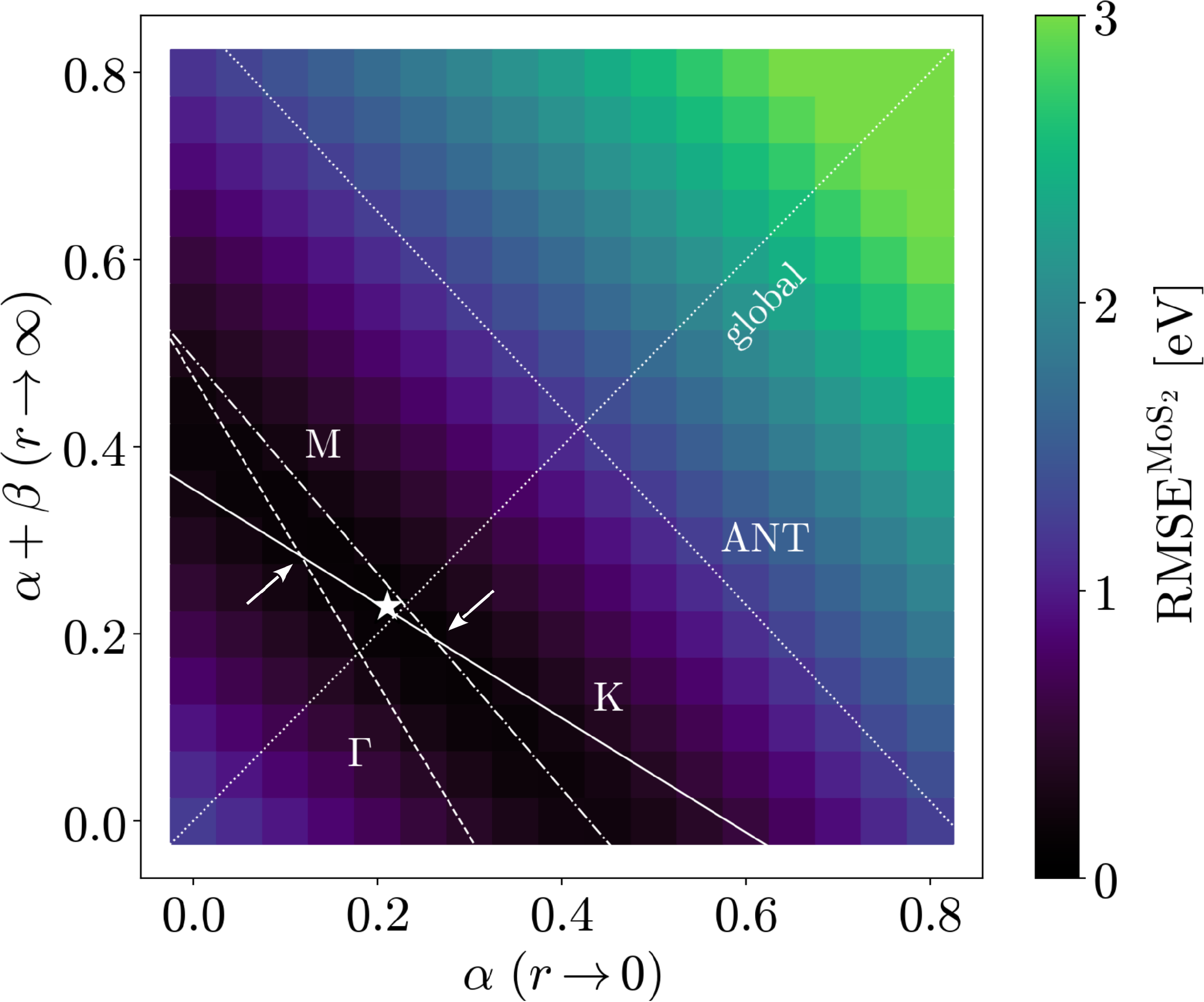}
    \caption{Root mean square error of the bandgap of isolated 1L-MoS$_2$ calculated with the CAM hybrid with respect to the $G_0W_0$@PBE benchmark calculation [RMSE$^\mathrm{MoS_2}$, Eq.~\eqref{eq:rmse_mos2}], as a function of the parameters $\alpha$ and $\beta$ for a fixed $\gamma=0.106\,a_\mathrm{B}^{-1}$. The white lines represent parameter values for which the CAM functional and $G_0W_0$@PBE reference yield identical gaps, resolved by high-symmetry point. RMSE$^\mathrm{MoS_2}$ is minimized for the parameters marked by the star. The arrows highlight the $\Gamma$-K ($\beta > 0$) and M-K ($\beta < 0$) intersections. The line labeled ``ANT" represents parameters for which CAM reproduces the benchmark HOMO-LUMO gap from $\Delta$PBE+PCM. PBE, PBE0, and HSE with default parametrizations correspond to $(\alpha,\beta) = (0, 0)$, $(\alpha,\beta) = (0.25, 0)$, and $(\alpha,\beta) = (0.25, -0.25)$, respectively.}
    \label{fig:rmse_mos2}
\end{figure}

The analysis reported above shows the known pitfalls of PBE to reproduce the band structure of 1L-MoS$_2$: not only does it feature a substantial underestimation of the bandgap, but the band dispersion in particular between M and K departs considerably from the reference including the QP correction.
In the following, we investigate whether appropriately parametrized CAM functionals overcome the issues of PBE.
To do so, we tune the parameters $\alpha$, $\beta$, and $\gamma$ [Eq.~\eqref{eq:cam}] to reproduce the $G_0W_0$@PBE band structure of MoS$_2$. As an error metric, we employ the root-mean-squared error (RMSE) of the band structure obtained with CAM functional with respect to the $GW$ reference:
\begin{align}\label{eq:rmse_mos2}
 &\mathrm{RMSE}^\mathrm{MoS_2}(\alpha,\beta,\gamma) = \bigl\{\nonumber\\ &\hspace{1.1cm}(1/5)\left[E_{g\Gamma}^\mathrm{CAM}(\alpha, \beta, \gamma) - E_{g\Gamma}^{GW}\right]^2
 \nonumber\\ &\hspace{0.65cm}+(2/5)\left[E_{g\mathrm{M}}^\mathrm{CAM}(\alpha, \beta, \gamma) - E_{g\mathrm{M}}^{GW}\right]^2\nonumber\\
 &\hspace{0.65cm}+(2/5)\left[E_{g\mathrm{K}}^\mathrm{CAM}(\alpha, \beta, \gamma) - E_{g\mathrm{K}}^{GW}\right]^2\nonumber\\
 &\bigl\}^{1/2}.
\end{align}
The bandgap $E_{gX}$ is averaged over the high-symmetry points $X\in\{\Gamma, \mathrm{M},\mathrm{K}\}$, attributing twice the weight to M and K due to the existence of two corresponding inequivalent points in the Brillouin zone of MoS$_2$. The RMSE is monitored in parallel with the individual errors at $\Gamma$, M and K. For each $\gamma$ and high-symmetry point, there exists a straight line in the $(\alpha,\beta)$ plane along which the corresponding error vanishes, \textit{i.e.}, the CAM functional yields the same gap as $G_0W_0$@PBE (Fig.~\ref{fig:rmse_mos2}).

The best global hybrid ($\beta = 0$ and/or $\gamma = 0$) has an exact-exchange fraction of $\alpha = 0.22$, close to PBE0 in its standard parametrization ($\alpha=0.25$)~\cite{pbe0}, with an RMSE of about 0.13~eV. This is a fair improvement over rigidly scissors-shifted PBE, for which the RMSE amounts to 0.24~eV.
Introducing range separation and fixing $\gamma = 0.106\,a_\mathrm{B}^{-1}$ for illustration, the minimum moves slightly to $\alpha=0.21$, $\beta=0.02$, as marked by the star in Fig.~\ref{fig:rmse_mos2}; however, the corresponding decrease of RMSE is negligible. Screening the minimum RMSE over $\gamma = n\times 0.053~a_\mathrm{B}^{-1}$, $n\in\{0,1,2,3,4,10\}$, no substantial improvement is found, indicating that range separation is generally unable to substantially enhance the band structure (Sec.~S5).

Range separation does open up some possibilities if considering error metrics other than Eq.~\eqref{eq:rmse_mos2} to optimize the band structure locally instead of globally throughout the Brillouin zone. It can be of interest to produce the correct dispersion around the band extrema at K, \textit{e.g.}, in the calculation effective masses. This can be approximately achieved by simultaneously minimizing the bandgap error at both M and K, which is only possible with $\gamma>0$. To see why, we note that the intersection of the optimal-M and -K lines with the $\beta=0$ diagonal are independent of $\gamma$ and thus can be seen as pivots around which the lines revolve upon variation of $\gamma$ (Figs.~\ref{fig:rmse_mos2} and S3). For $\gamma=0$, the optimal-M and -K lines run vertically and in parallel at $\alpha=0.23$ and $0.22$, respectively. With range separation, the lines tilt to different degrees and cross at some point in the off-diagonal region $\beta < 0$ (right arrow in Fig.~\ref{fig:rmse_mos2}), allowing for simultaneous reproduction of the correct gap at M and K. Here, we adjust $\gamma$ until the M-K intersection falls onto the $\alpha +\beta = 0$ axis, yielding an optimal HSE-like functional for $\alpha=0.28$ and $\gamma = 0.034\,a_\mathrm{B}^{-1}$, producing the CAM band structure featured in Fig.~\ref{fig:unit_bands}. Obviously, minimizing the errors at M and K in this manner entails a higher error at $\Gamma$, since the M-K crossing point does not represent the minimal RMSE according to Eq.~\eqref{eq:rmse_mos2}, which is always close to $(\alpha,\beta)=(0.22, 0)$. The increased error can be traced back mainly to an overestimation of the lowest conduction states at $\Gamma$ by about 0.5~eV, while the valence band is less affected. 

CAM functionals in the time-dependent extension of DFT have furthermore been established as a valid alternative to the expensive Bethe-Salpter formalism for the calculation of optical properties, both for molecules~\cite{kron+2012jctc, refa+2013prb} and two-dimensional materials~\cite{huang+2017jpcc, rama+2019prm}. In the present case of 1L-MoS$_2$, it is of interest to optimize the gaps at K and $\Gamma$ for such an application, since optical transitions at those points dominate the absorption onset~\cite{qiu2016prb}. Also this objective cannot be achieved with a global hybrid functional, since the optimal $\Gamma$ and K lines run vertically at $\alpha=0.18$ and $0.22$, respectively. Choosing $\gamma>0$ and large enough, the lines start to intersect in the off-diagonal region $\beta > 0$ (left arrow in Fig.~\ref{fig:rmse_mos2}). Adjusting $\gamma$ until the crossing point falls onto the $\alpha+\beta=1$ axis, we obtain values of $\alpha = 0.02$ and $\gamma = 0.032\,a_\mathrm{B}^{-1}$. In Ref.~\citenum{rama+2019prm}, optimal parameters of $\alpha=0.11$ and $\gamma=0.020\,a_\mathrm{B}^{-1}$ were determined; the same constraint ($\alpha+\beta=1$) was enforced based on arguments regarding the nature of screening in two-dimensional materials. However, a different cost function was employed, optimizing the gap at K but not at $\Gamma$. This likely explains the reported overestimation of excitation energies pertinent to transitions occurring at the center of the Brillouin zone.

Independent of range separation, we find that hybrid functionals producing accurate bandgaps systematically underestimate absolute QP energies. To encounter this issue, we tested also the possibility to add the ionization potential as a feature in the optimization procedure (Sec. S5). However, an improvement in this regard comes at the cost of worsening the values for the bandgaps.

\subsubsection{Substrate-induced bandgap renormalization}

\begin{figure}[h]
    \centering
    \includegraphics[width=0.46\textwidth]{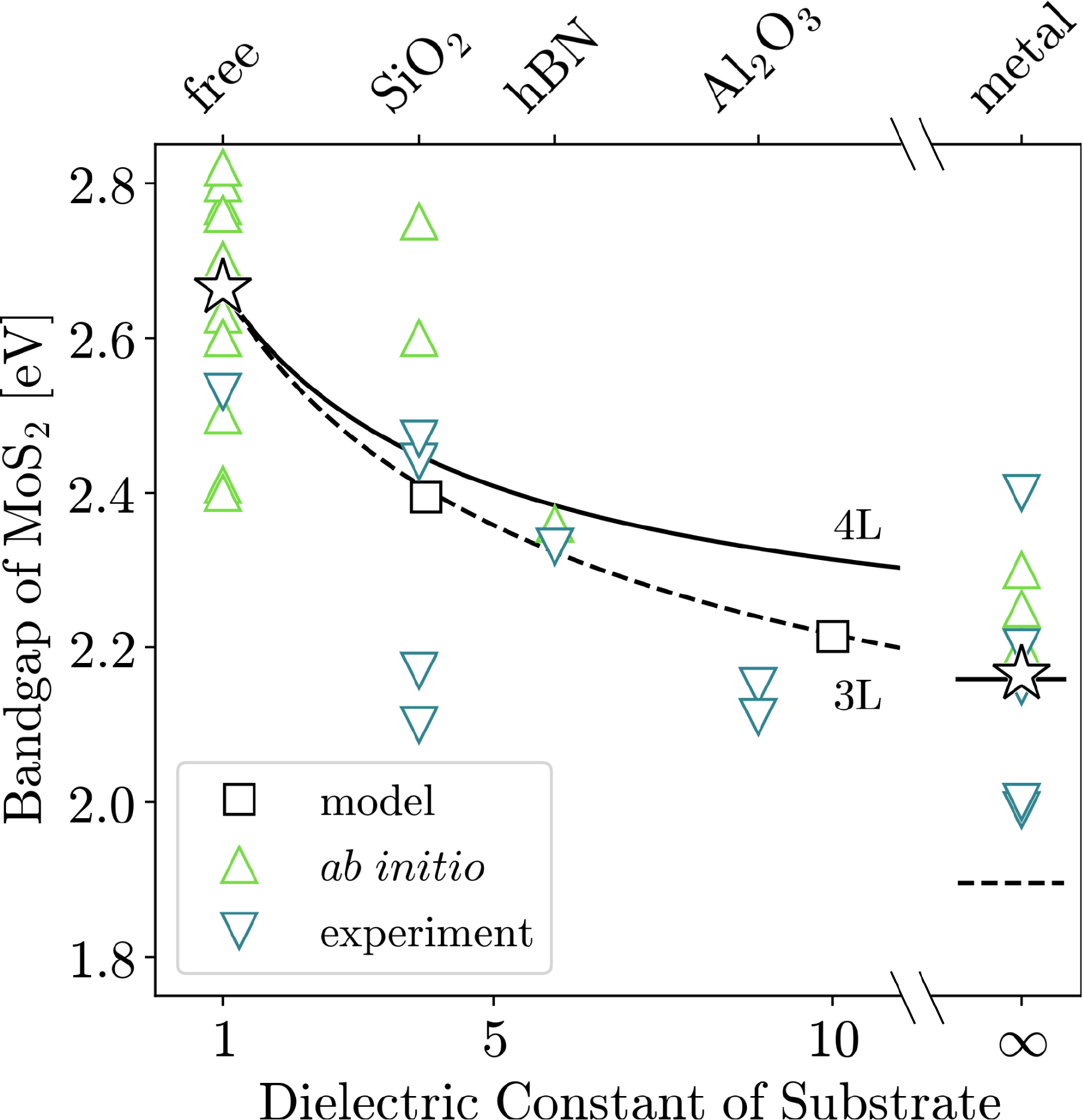}
    \caption{Bandgaps of MoS$_2$ as a function of the substrate dielectric constant $\varepsilon_s$. The stars mark the average of all values at a given $\varepsilon_s$. The lines indicate the screened bandgap as predicted by the 3-layer (3L, dashed) and 4-layer (4L, solid) models with respect to the free-layer average. Under the ``metal" substrate, we feature results for graphite, gold, and silver. The square data points (``model") are from the 3L model of Ref.~\citenum{cho+berk2018prb}. Theoretical gaps are taken from Refs.~\citenum{ryou+2016sr, utam+2019ne, drue2017natcom, naik+jain2018prm, qiu+2015prl, qiu2016prb, sokl+2014apl, koms+kras2012prb, moli+2013prb, shi+2013prb, huse+2013prb, rama2012prb, conl+2013nl, chei+lamb2012prb, liang+2013apl}; experimental ones from Refs.~\citenum{shi+2015am, park+2021acsn, murr+2019prb, shi+2016ami, hill+2015nl, huan+2015natcom, zhan+2014nl, chiu+2015natcom, rigo+2016prb, klei+2019apl};  dielectric constants from Refs.~\citenum{geic+1966pr, Grove1967PhysicsAT, harm+1994jap}.}
    \label{fig:bandgap_substrate}
\end{figure}

HSE \textit{in its standard parametrization} is reputed for its capability of predicting the fundamental gap of bulk semiconductors~\cite{jane+2009pccp, paie+2006jcp, bros2008jcp}. However, the gap it offers for MoS$_2$ (2.16~eV) is far smaller than that from $GW$, which is around 2.7~eV, as seen above. The bandgap of 1L-MoS$_2$ is large relative to its bulk counterpart, which can be attributed to the less effective screening in the two-dimensional limit, leading to stronger exchange interactions. Notice that HSE was optimized for the isotropic screening of typical bulk materials~\cite{hse03,hse06}. Now, it is worth specifying that free-standing 1L-MoS$_2$ is rarely considered in experiments: usually, the material is placed on top of a substrate. In this case, the system experiences a screening that is intermediate between the two-dimensional configuration and the three-dimensional one.
Obviously, this type of screening depends strongly on the type of substrate, particularly on its dielectric constant. For this reason, the concept of the QP gap as a material constant is less appropriate, and a wide range of corresponding values has been reported for 1L.MoS$_2$ (Fig.~\ref{fig:bandgap_substrate}). Uncertainties are exacerbated by the challenging convergence of $GW$ calculations for two-dimensional materials~\cite{qiu2016prb}, especially upon (approximate) inclusion of substrates, as well as by variables related to the adopted measurement setup. For example, bandgap values reported from scanning tunneling microscopy are significantly different from those given by optical absorption measurements on identical samples~\cite{klei+2019apl}. 

To better rationalize the effect of the substrate, we compute the correction to the bandgap as described in Sec.~\ref{sec:model}. In Ref.~\citenum{cho+berk2018prb}, the
scenario of a TMDC monolayer on a substrate was modeled as a dielectric slab surrounded by a semi-infinite dielectric medium underneath and vacuum above; $W$ and $W_0$ were calculated through the method of recursive image charges~\cite{kuma+taka1989prb} with parameters such as the slab thickness chosen somewhat arbitrarily. Here, we use a well-defined method for determining the thickness from first principles, finding 5.35~\AA\,\,(Sec.~S3). Since this value is significantly smaller than the interlayer separation in bulk MoS$_2$, which we estimated as $\sim$6.3~\AA, it stands to reason that the adsorbed 1L-MoS$_2$ modeled as a dielectric continuum should be separated from the substrate by a vacuum spacer of about $0.95$~\AA. The corresponding Poisson problem can be solved with partial Fourier transforms, an approach that becomes more convenient than the image-charge method with increasing numbers of dielectric layers (Sec.~S6). Contrasting the three-layer (3L) and four-layer (4L) models without and with the vacuum spacer, respectively, we find neither universally superior to the other (Fig.~\ref{fig:bandgap_substrate}). The high variance in the canvassed data for dielectric substrates - \textit{e.g.}, in the case of SiO$_2$ - renders the corresponding results from experiments and MBPT calculations an unreliable benchmark. However, 4L appears to improve significantly the asymptotic value of the bandgap (2.16~eV) assumed for metallic substrates ($\varepsilon\rightarrow\infty$), which is in good agreement with the more consensual average of reported bandgaps for such systems (2.17~eV). 

The bandgap of 1L-MoS$_2$ on a metal is close to the value obtained with default HSE (2.16~eV). This finding can be rationalized as follows. In the case of adsorption on a metallic substrate, the Coulomb interaction is only screened in the region below the monolayer, but by a highly polarizable material. This has roughly the same effect as the isotropic \textit{average} dielectric that HSE implicitly assumes. Thus, one could say that this functional roughly models the system not in vacuum, but rather atop a metallic substrate, though obviously missing all TMDC-substrate interactions beyond the electrostatic gap renormalization.

\subsubsection{HOMO-LUMO gap of ANT: PBE and $\Delta$PBE}\label{sec:ant_dscf}

In the next step of our analysis, we inspect the electronic structure of the organic component of the considered hybrid interface, namely the ANT molecule [Fig.~\ref{fig:unit_geom}b)].
Also in this case, the KS energy levels in the PBE approximation of $V_{xc}$ do not provide a good estimate of the fundamental gap.
The KS gap with PBE amounts to 2.37~eV, which is a most severe underestimation when compared to the value of 6.44~eV obtained with the same functional in the  $\Delta$SCF approach. This scheme is known to produce results on par with $GW$ for finite systems, where the Hartree term is the dominant driver of electronic relaxation~\cite{godb+whit1998prl, mart+2016book}. However, the value of 6.44~eV obtained herein for ANT still undershoots measured results by $\sim$0.5~eV~\cite{cocc+2023jpm}.

In Ref.~\citenum{cocc+2023jpm}, it is also shown that the agreement between $\Delta$SCF and experiment becomes excellent for acenes in dielectric environments if polarization is self-consistently included through the PCM. For the ANT@MoS$_2$ hybrid system (Fig.~\ref{fig:geometry}), which will be considered in detail in Sec.~\ref{sec:hybrid}, the gap between the lowest unoccupied molecular orbital (LUMO) and the highest occupied molecular orbital (HOMO) is reduced to 4.66~eV when modelling the TMDC as a dielectric slab. This renormalized gap should be taken as a benchmark value for tuning the CAM functional~\cite{kron+kuem2018am}. Thus, the ANT target gap includes an aspect of the interaction with 1L-MoS$_2$, namely the polarization-induced renormalization, but not \textit{vice versa}. This can be justified by arguing that the additional screening introduced by the spatially confined molecule is negligible compared to that occurring within 1L-MoS$_2$. The validity of this assumption will be confirmed below by the results of fully atomistic $G_0W_0$@PBE simulations of the hybrid system.

To quantify the influence of a substrate underneath the TMDC on the energy levels of ANT, we calculate the energy gap using the $\Delta$PBE+PCM approach with $\varepsilon_s\rightarrow \infty$, \textit{i.e.}, assuming a metallic substrate. We obtain a gap change of only $\sim$0.05~eV with respect to ANT adsorbed on free-standing MoS$_2$ ($\varepsilon_s = 1$), which is negligible compared to the change of 1.78~eV due to the TMDC itself. This can be explained by the high dielectric constant of MoS$_2$ ($\varepsilon>10$), making it already quite effective at screening. The situation would most likely be different for insulating (low-$\varepsilon$) two-dimensional materials such as boron nitride, where the additional substrate screening can be expected to play a more important role.

It is easy to adjust a CAM functional to correctly reproduce the electronic structure of the molecule, compared to an extended system. Lines marking the optimal parameters for the HOMO-LUMO gap and the ionization potential almost coincide for all $\gamma$ (Fig.~S3), opening up a myriad of valid choices of the parameters. This freedom has been exploited in the past to optimize the electronic structure beyond the fundamental gap, considering the energy gaps within the valence subspace~\cite{brom+2017jcp}. Here, we aim to understand whether this freedom can instead be used to yield a satisfactory description of the electron structure of hybrid inorganic-organic systems in an all-atomistic simulation.

\subsection{Hybrid systems}
\label{sec:hybrid}

\begin{figure}
    \centering
    \includegraphics[width=0.44\textwidth]{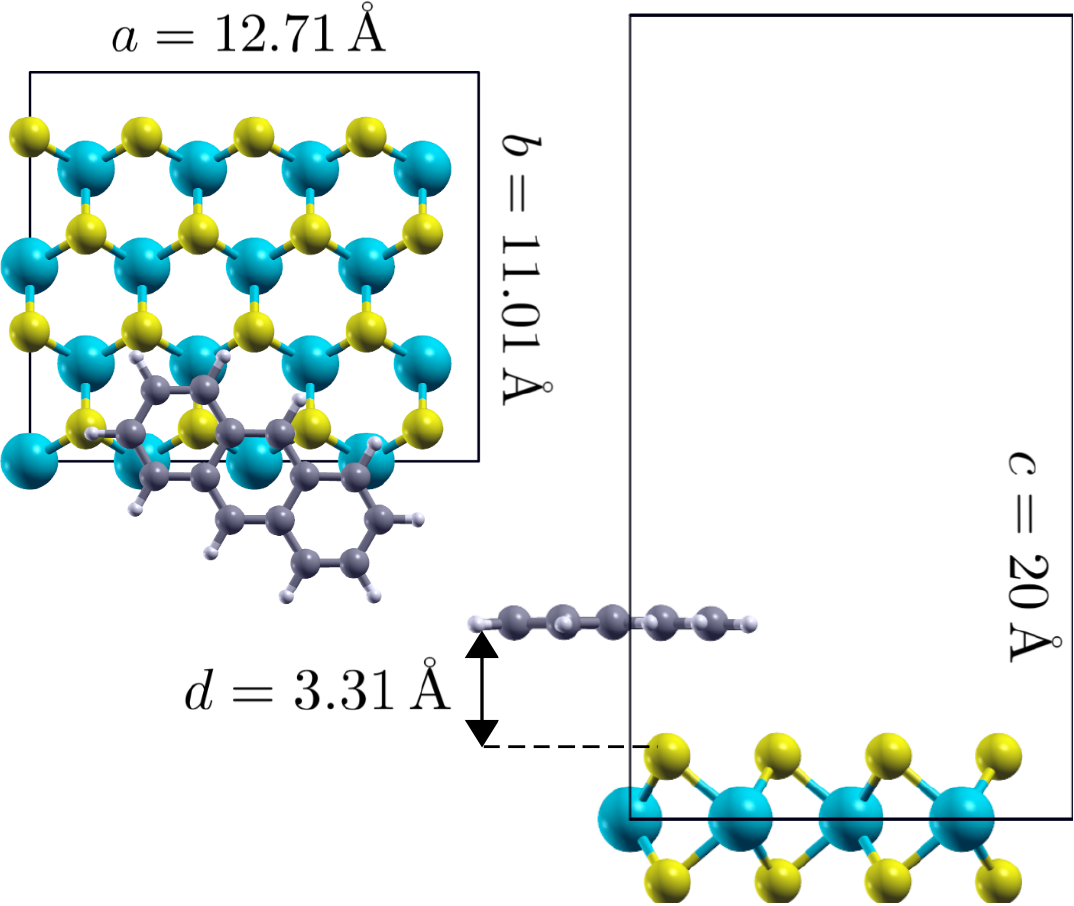}
    \caption{Unit cell of the hybrid system combining an anthracene molecule with a $4\times2\sqrt{3}$ supercell of MoS$_2$, visualized with XCrysDen~\cite{xcrysden}. Mo, S, C, and H atoms are depicted in turquoise, yellow, grey, and white, respectively.}
    \label{fig:geometry}
\end{figure}

Having considered the subsystems individually, we now turn to the analysis of the atomistically modeled ANT@MoS$_2$ hybrid interface (Fig.~\ref{fig:geometry}). We compare the results of semi-local PBE, $G_0W_0$@PBE, a general CAM functional, 
and the subsystem calculations of Sec.~\ref{sec:constituents}.

\subsubsection{PBE results}

\begin{figure}[t]
    \centering
    \includegraphics[width=0.49\textwidth]{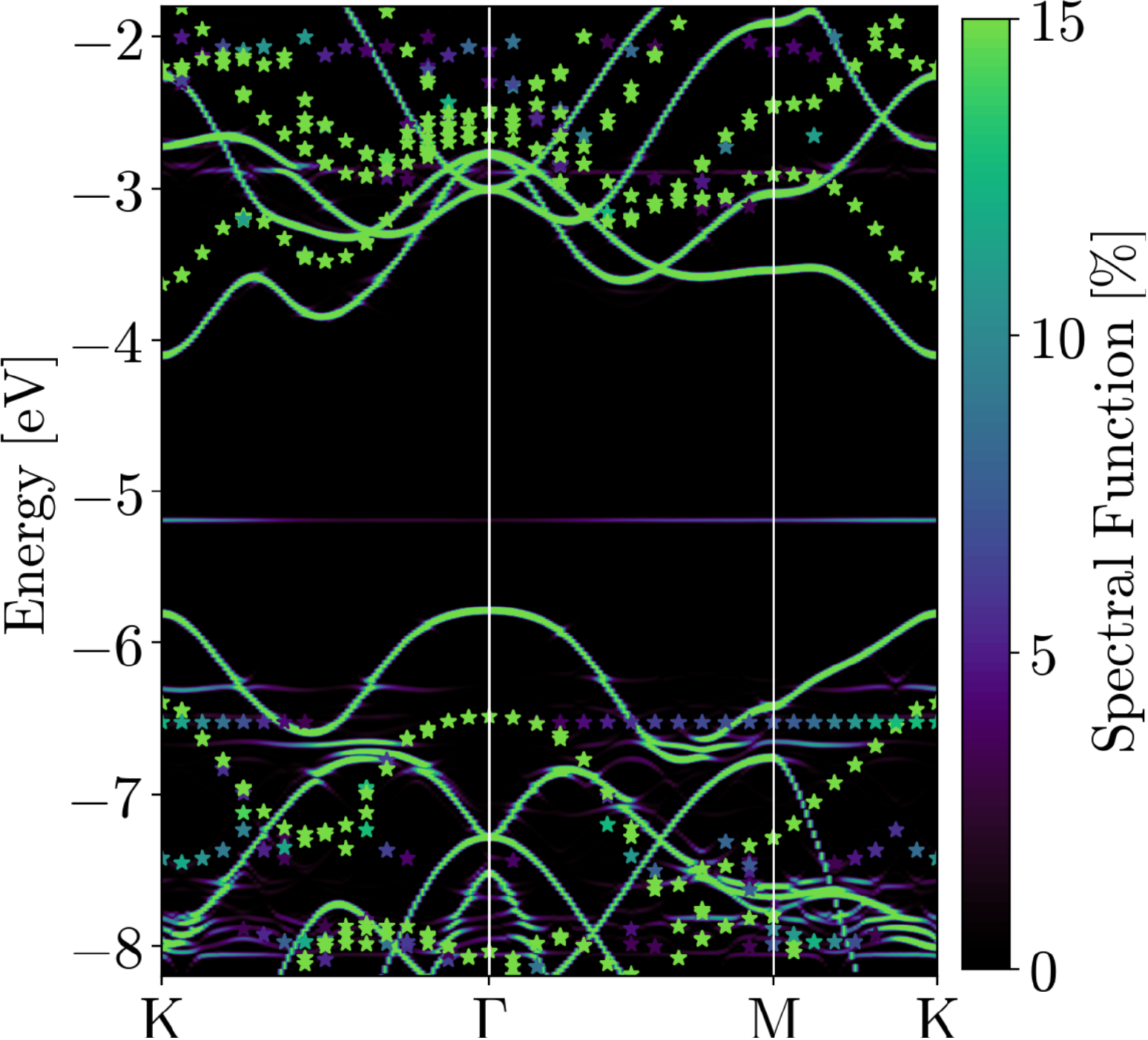}
    \caption{Unfolded band structure of the ANT@MoS$_2$ hybrid system. The colormap in the background represents the PBE result, while the scattered overlaid stars come from $G_0W_0$@PBE. 
    }
    \label{fig:ubs}
\end{figure}

In the band structure calculated with PBE and unfolded to the unit cell of MoS$_2$ (Fig.~\ref{fig:ubs}),
the TMDC states are found at energies about 0.15~eV
higher than in the free-standing monolayer. This can be explained in terms of electron transfer from ANT to MoS$_2$, stabilizing and destabilizing the states of ANT and MoS$_2$, respectively, through emerging partial charges. Similar trends have been found for other conjugated molecules adsorbed on TMDC monolayers~\cite{habi+2020ats, mela+22pccp}. The molecular orbitals appear as \textbf{k}-delocalized wavepackets since Bloch waves with different wavevectors have to be superposed to construct such localized states. The HOMO is found mid-gap at an energy of -5.19~eV, which is about 0.2~eV smaller than that of the isolated molecule; the difference is in equal parts due to electron donation to MoS$_2$ and residual interactions between ANT replicas in neighboring cells. For the same reasons, the LUMO at -2.9~eV is about 0.25~eV lower in energy compared to the gas phase counterpart. The gap of the molecule thus changes only by 0.05~eV due to adsorption on the substrate, reflecting the failure of 
current state-of-the-approximations $xc$ approximations to capture the image-charge induced renormalization of the order of 1~eV~\cite{neaton2006prl}. The band alignment is of staggered type.

\subsubsection{$GW$ results}

The application of the $G_0W_0$ correction to the PBE results leads to a substantially different picture (Fig.~\ref{fig:ubs}). The HOMO of ANT is significantly downshifted in energy and ends up below the VBM of MoS$_2$, thus turning the type-II level alignment predicted by PBE into a type-I lineup with the molecular gap encompassing the TMDC one. Generally, occupied ANT states are downshifted by an additional $\sim$1~eV compared to MoS$_2$ counterparts. The renormalized HOMO-LUMO gap from $G_0W_0$@PBE agrees within $0.1$~eV with the one predicted by a $\Delta$SCF+PCM calculation in which the molecule is modeled explicitly and the substrate implicitly (Sec.~\ref{sec:ant_dscf}). The bandgap of MoS$_2$ remains unchanged with respect to the one of the free-standing monolayer. This supports our previous claim that the additional screening due to the (nearly isolated) molecule is negligible for QPs inside the TMDC. On a computational side note, it is found that, like in the pristine TMDC, the absolute QP energies of the frontier states as obtained by the complete-basis extrapolation can be reproduced fairly well by centering the MoS$_2$ bandgap from an under-converged $G_0W_0$@PBE calculation around the MoS$_2$ mid-gap energy of the PBE starting point, ignoring the molecular states~(Fig.~\ref{fig:alignment}). This offers a cheap alternative to the extrapolation technique, requiring only a single $G_0W_0$ calculation.

While the updated level alignment given by the $G_0W_0$ correction certainly represents an improvement over PBE, the method comes with its own set of issues. Orbitals are not updated as the self-energy is not diagonalized; only diagonal elements are computed. States mixed in the PBE electronic structure remain so even though the energies do not align anymore; conversely, no new interaction is introduced where the corrected alignment should lead to such. Furthermore, as the QP corrections are very different for ANT and MoS$_2$ states, hybridized inorganic-organic states stemming from the PBE calculation are corrected by an unphysical amount. This is most obvious for the HOMO-1 of the molecule, which is visible at K around -6.3 and -7.4~eV in the PBE and $G_0W_0$@PBE electronic structures, respectively (see Fig.~\ref{fig:ubs}). This orbital is increasingly hybridized with MoS$_2$ states upon following the path from K towards $\Gamma$, leading to a continuously decreasing QP correction. This turns the energetically flat molecular state obtained from PBE into an unphysically dispersive one, peaking at the intersection with the MoS$_2$ band, where the admixture of TMDC states is presumably largest. Also the LUMO can be found to be widely scattered in energy around -2~eV after the QP correction, reflecting the varying degree of MoS$_2$ character in the molecule-TMDC hybridized states throughout the Brillouin zone. The issues of hybridization in the electronic structure given by PBE has been touched upon in Ref.~\citenum{aden+liu2021jcp}, albeit not clearly pointed out as an artifact of the method. It represents a violation of the fundamental requirement for the applicability of $G_0W_0$ that the KS orbitals be reasonable approximations to the QP orbitals. Due to the poor description of mixed states, it seems unlikely that a $G_0W_0$@PBE electronic structure provides a reliable starting point for the calculation of optical properties. While, in principle, it should be possible to remedy this deficiency by updating the Green's function, it is questionable whether the results of this procedure are able to justify the considerable computational costs expected for self-consistent $GW$ calculations on such hybrid interfaces.

\subsubsection{Results from the CAM functional}

In principle, one should optimize the parameters by defining an error metric featuring relevant subsystem quantities such as the fundamental gap and the ionization potential of both MoS$_2$ and ANT in order to have a predictive CAM functional. However, comparing the zero-error lines for the two subsystems (Fig.~\ref{fig:rmse_mos2}) foreshadows that it is a difficult task to find parameters yielding satisfactory results for both subsystems. Here, we pursue a different route that allows us to make more general statements independent of the specific choice of the error metric. For a CAM hybrid to be a worthwhile alternative to $GW$, it should similarly be able to correct the major shortcomings of the PBE result. We identify the qualitatively wrong alignment predicted by PBE as the principal issue, as it renders the entire band structure essentially useless. Remedying this problem is the minimum requirement for CAM functionals. 

\begin{figure*}
    \centering
    \includegraphics[width=0.95\textwidth]{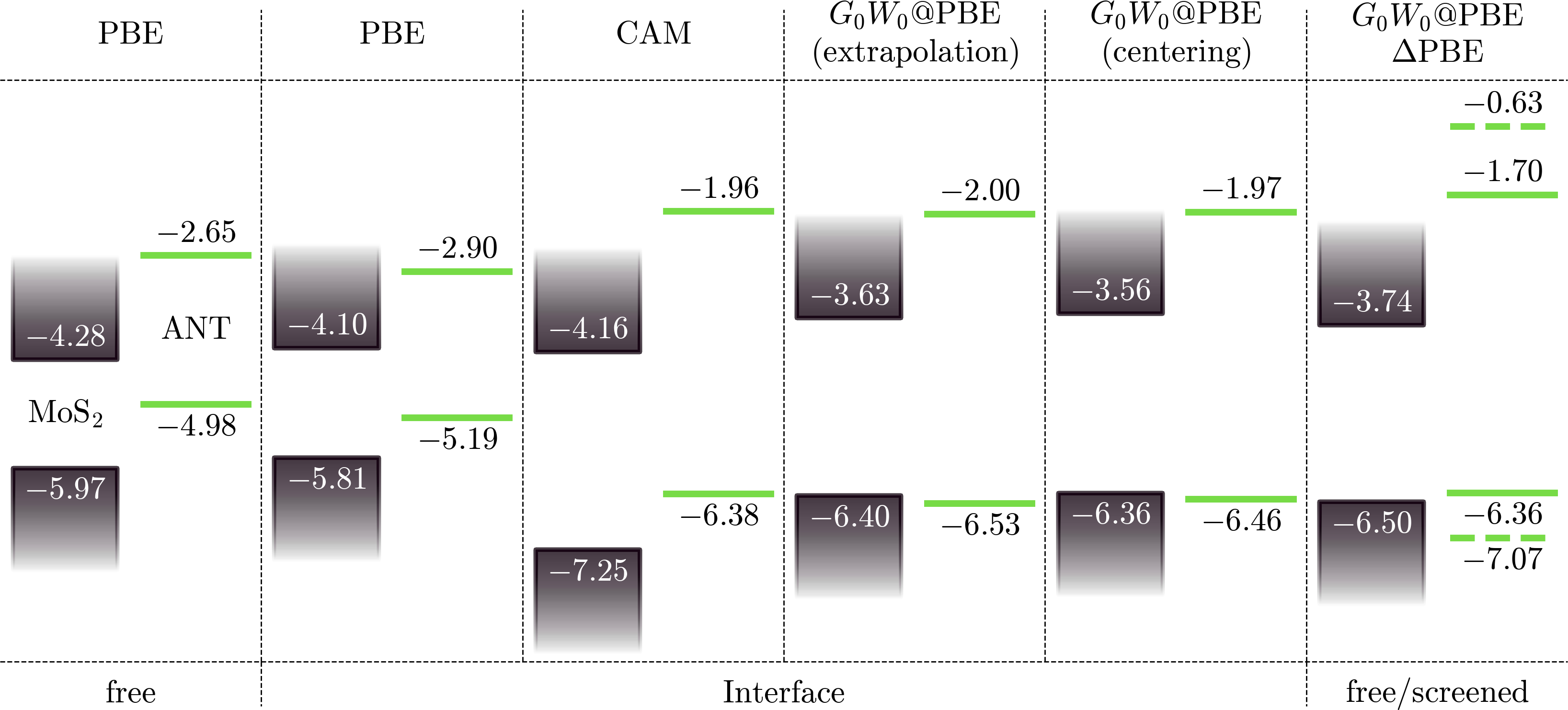}
    \caption{Band alignment at the ANT@MoS$_2$ interface, including both isolated (``free") systems and interfaces. For the $G_0W_0$@PBE calculation of the hybrid interface, complete-basis extrapolation and the PBE/$GW$ gap-centering method for obtaining absolute band energies are compared. The CAM result uses parameters $\alpha = -\beta = 1$ and $\gamma=0.145\,a_\mathrm{B}^{-1}$, which have been found to optimize the squared gap error averaged over MoS$_2$ and ANT. The rightmost column combines a $G_0W_0$@PBE calculation of MoS$_2$ in its unit cell with a $\Delta$SCF(+PCM) calculation of the molecule. For the latter, both gas-phase (dashed) and screening-renormalized (solid) are shown. All energies are expressed in eV.}
    \label{fig:alignment}
\end{figure*}

To see whether it is able to accomplish that, we monitor directly the energetic separation between the HOMO and the VBM. An improvement over PBE would correspond to a reduction of this separation, bringing the system closer to the type-I level alignment predicted by $G_0W_0$@PBE. However, this is not what we observe. The HOMO-VBM separation is rather insensitive to adjustments of all three CAM parameters. The small changes that do occur are furthermore in the wrong direction: the inclusion of Fock exchange consistently \textit{increases} the HOMO-VBM gap (Fig.~\ref{fig:alignment}). From this, we conclude that range-separated hybrid functionals are inappropriate for the simulation of interfaces between TMDCs and molecules and do not necessarily represent an improvement over DFT with semi-local $xc$ functionals. Furthermore, it shows that the hybridization issues found for $G_0W_0$@PBE cannot be cured by using a hybrid functional as the DFT starting point, which is often seen as an alternative to self-consistent $GW$.

\section{Discussion}
\label{sec:discussion}

It could be argued that the example of ANT@MoS$_2$ is a worst-case scenario for DFT and $G_0W_0$@DFT: the molecule is small and thus has a large HOMO-LUMO gap. As the underestimation of the fundamental gap by DFT is percentual, it is very large in absolute terms for ANT when compared to MoS$_2$, which has a much smaller gap to begin with. Hence, absolute orbital energies cannot be compared well across the organic and inorganic components, leading to a poor description of the interfacial electronic structure by PBE. It also leads to the discrepancies in the size of the QP corrections that is the main root of artifacts in the $G_0W_0$ results. 

For larger molecules, more accurate predictions can be expected, due to the smaller HOMO-LUMO gap, both for a $G_0W_0$-corrected and a pure DFT result. However, in this case, either method has its own issues: while it is possible to compare the QP corrections for molecular and TMDC states and thus judge whether large differences, which are the root of hybridization artifacts, exist, the $G_0W_0$ approach suffers from poor scalability with respect to system size. Meanwhile, DFT is comparatively efficient and thus, can technically be employed for large adsorbed molecules, but there is no immediate indication of the quality of the results since the size of the bandgap error is not known exactly.

At last, some final remarks about the band alignment given by different methods (Fig.~\ref{fig:alignment}). PBE predicts the molecular orbitals to be reduced in energy upon deposition on a TMDC. As mentioned, this is caused by intermolecular interactions and charge transfer. The amount of charge transfer can be assumed to be overestimated since the HOMO is too high in energy and thus too close to the CBm when compared to higher-order theory such as MBPT. In addition, the lack of a derivative discontinuity in PBE leads to an overestimation of the electron affinity of MoS$_2$, reinforcing this trend. Both of these shifts are also present in the $G_0W_0$@PBE result. Without them, the band alignment would remain of type II, with the HOMO and valence band of the TMDC almost resonant. In fact, the energies of the HOMO and the LUMO agree quite well between $\Delta$PBE+PCM and $G_0W_0$@PBE when the shift of 0.2~eV due to (overestimated) charge transfer and intermolecular interactions is subtracted. 

The accuracy of the alignment given by the subsystem calculations as well as their low computational costs suggests working with them directly and forgoing the ordeal of an MBPT simulation of the whole hybrid system, which anyway is plagued by the artifacts of the DFT starting point. Residual effects such as charge transfer leading to readjustments of the levels can potentially be included in a perturbative fashion; predictions about orbital-specific interactions and hybridization can be made on the basis of symmetry analysis and atomic orbital decomposition~\cite{krum-cocc21es}.

\section{Summary and Outlook}\label{sec:conclu}

In summary, we presented a comprehensive methodological analysis of \textit{ab initio} approaches for simulating the electronic structure of low-dimensional inorganic/organic interfaces, taking as an example an anthracene molecule adsorbed on a single layer of MoS$_2$. Methods surveyed are density functional theory with both pure and flexible hybrid approximations for exchange and correlation effects, many-body perturbation theory in the flavor of $G_0W_0$, and implicit models with parameters determined from first principles. While generally unable to achieve quantitative agreement with results obtained with $GW$, hybrid functionals are found to be generally superior to the semi-local ones in the description of the isolated subsystems, with range separation opening up additional possibilities to tune the functional, \textit{e.g.}, for the calculation of optical properties. The widespread PBE0 and HSE functionals give reasonable results for band structure and bandgap; the former functional models free-standing 1L-MoS$_2$, while the latter can be viewed as implicitly including a screening substrate renormalizing the bandgap. We investigated this renormalization effect with an electrostatic model, yielding results in satisfactory agreement with experimental data. 

For the inorganic/organic mixed-dimensional heterostructures, the verdict about hybrid functionals is less favorable. By comparison to $GW$ results, we found that semi-local functionals give qualitatively wrong results for interfacial properties, such as level alignment and electronic hybridization, which hybrid functionals are unable to systematically improve. Thus, they also trouble single-shot $G_0W_0$ calculations by providing poor starting points. Consequently, results of such expensive simulations, while undoubtedly superior to the ones obtained with DFT, are also flawed and do not constitute a comprehensive picture of the electronic structure, making it questionable to further process them, \textit{e.g.}, for the determination of optical properties. The main issue in this regard arises from hybridized states in the underlying electronic structure, which obtain inaccurate quasiparticle corrections that erroneously distribute molecular states over a wide energy range. 

We conclude that \textit{ab initio}-parametrized implicit models are serious competitors for full-fledged atomistic many-body calculations, yielding energy levels generally within 0.1~eV of corresponding results from $G_0W_0$@PBE, all while being orders of magnitude cheaper in terms of computational complexity. In the future, we aim to build upon this and ameliorate the accurate first approximation to the interfacial electronic structure from subsystem \textit{ab initio} calculations and model interactions by including coupling terms beyond the dominant polarization-induced level renormalization.

\subsection*{Acknowledgements}
This work was supported by the German Research Foundation through the Collaborative Research Center HIOS (Project number 182087777 - SFB 951). CC acknowledges additional funding from the German Federal Ministry of Education and Research (Professorinnenprogramm III), and from the State of Lower Saxony (Professorinnen für Niedersachsen). The computational resources were provided by the North-German Supercomputing Alliance (HLRN). 

\subsection*{Conflict of interest}
The authors declare no conflict of interest.

\subsection*{Data availability statement}
The data that support the findings of this study are openly available in Zenodo at DOI 10.5281/zenodo.7620432


%

\end{document}